# NASA ExoPAG Study Analysis Group #5: Flagship Exoplanet Imaging Mission Science Goals and Requirements Report

T. Greene[*] and C. Noecker[†] for the ExoPAG SAG #5 Team
25 March 2013

## Abstract

The NASA Exoplanet Program Analysis Group (ExoPAG) has undertaken an effort to define mission Level 1 requirements for exoplanet direct detection missions at a range of sizes. This report outlines the science goals and requirements for the next exoplanet flagship imaging and spectroscopy mission as determined by the flagship mission Study Analysis Group (SAG) of the NASA Exoplanet Program Analysis Group (ExoPAG). We expect that these goals and requirements will be used to evaluate specific architectures for a future flagship exoplanet imaging and spectroscopy mission, and we expect this effort to serve as a guide and template for similar goals and requirements for smaller missions, an effort that we expect will begin soon. These goals and requirements were discussed, determined, and documented over a 1 year period with contributions from approximately 60 volunteer exoplanet scientists, technologists, and engineers. Numerous teleconferences, emails, and several in-person meetings were conducted to progress on this task, resulting in creating and improving drafts of mission science goals and requirements. That work has been documented in this report as a set of science goals, more detailed objectives, and specific requirements with deliberate flow-down and linkage between each of these sets. The specific requirements have been developed in two categories: "Musts" are non-negotiable hard requirements, while "Discriminator" requirements assign value to performance in areas beyond the floor values set by the "Musts." We believe that this framework and content will ensure that this report will be valuable when applied to future mission evaluation activities. We envision that any future exoplanet imaging flagship mission must also be capable of conducting a broad range of other observational astrophysics. We do not set requirements for this other science in this report but expect that this will be done by the NASA Cosmic Origins Program Analysis Group (COPAG).

## 1 Introduction

In February 2011 a single study analysis group (SAG) of the NASA Exoplanet Program Analysis Group (ExoPAG) was created to engage the scientific community in outlining the science goals and requirements for the next exoplanet flagship imaging and spectroscopy mission. By this time the Exoplanet Exploration Program was following NASA direction to reduce investment in infrared interferometry as a possible architecture for this mission, and instead, focus on single-aperture visible telescopes with internal coronagraphs or external starshades.[‡] ExoPAG SAG #5 was tasked with defining the science goals and requirements for a flagship imaging mission in the 2020 decade in a way that was independent of specific mission architectures, although, for example, we expect that a telescope aperture of at least 4 m will be required. The NASA astrophysics community also expects the next exoplanet flagship mission to serve jointly as the flagship mission for NASA optical and UV astrophysics as suggested in the New Worlds, New Horizons Decadal Survey report. The ExoPAG and the Cosmic Origins Program Analysis Group (COPAG) have endorsed this notion, and the COPAG has agreed to develop the non-exoplanet requirements for this mission.

---

[*] NASA Ames Research Center, Tom.Greene@nasa.gov
[†] NASA Jet Propulsion Laboratory, California Institute of Technology, Charley.Noecker@jpl.nasa.gov
[‡] Infrared interferometry is still considered a viable future technology for characterizing exoplanets.[1] And the suite of atmospheric biomarker gases that might be detected at thermal-IR wavelengths is complementary to those in the visible and near-IR; and so ultimately any potentially habitable planet that is found should be studied in both wavelength ranges.[2] But NASA concluded in 2004 that a visible/near-IR direct imaging mission is probably easier and cheaper, and should be first.[3]

## 1.1  Scope of this Report

This document outlines the comprehensive science goals, more detailed objectives, and initial Level 1 science and mission requirements for the next NASA flagship exoplanet mission as determined by this exoplanet flagship SAG. The Science Goals are general statements of what science is intended to be achieved by this mission. These are made more specific in the derived list of Objectives, and then even more specific in the list of requirements. The Science Goals and Objectives can be considered Level 0 and 0.5 descriptions that define the Level 1 requirements.

The work done for this report exploits and builds upon the significant amount of work done over the past decade to define science goals, requirements, and mission architectures for future exoplanet imaging missions. We have particularly leveraged the Terrestrial Planet Finder-Coronagraph (TPF-C) Science and Technology Definition Team (STDT) report,[4] completed in 2006. The ExoPAG document "Points of Scientific Agreement" was drafted shortly after the January 2011 ExoPAG meeting and served as a starting point for defining what exoplanet characteristics should be characterized (atmospheric spectral features, orbit, mass). That document was also used to develop the highest level mission statement and scientific goals for this report.

We expect that a mission concept capable of achieving these goals – as well as significant other astrophysics ones – will be documented and presented to the 2020 Astronomy and Astrophysics Decadal Survey. There are no goals, objectives, or Level 1 requirements for any astrophysics fields beyond exoplanets included in this document; we expect that the COPAG will provide those at a later date.

## 1.2  Processes and People

Many people throughout the greater exoplanet science and technology communities contributed to the work in this report. Participants were invited to join at the January and June 2011 ExoPAG meetings and were also solicited by the Exoplanet Program office via email distribution in February 2011 and via the ExoPAG web site. We had preliminary discussions via email and one teleconference in May 2011 before deciding on an approach for the task at the June 2011 ExoPAG meeting in Alexandria, VA. There we decided to adopt a hierarchical set of science goals, science objectives, and requirements with clear flow-down and linkage between these elements.

We also decided then to adopt a two tiered Level 1 requirements structure, with a minimal set of firm requirements that must be met ("Musts" in our parlance) and a set of "Discriminator" requirements that assign value to improving performance beyond or outside of the Must requirement values. This structure was adopted to enable quantitative scoring of competing mission architectures (e.g., coronagraphs and starshades) using Kepner-Tregoe methods.[5,6] Eventually, weights will be assigned to Discriminators according to their scientific, technical, or programmatic importance. In the present work Musts and Discriminators were selected to be specific enough that they correspond to concrete figures of merit. We have identified Discriminators but did not assign weights to them, because the flagship mission is still far in the future; scientific and technical progress before its launch will change the scientific values of any weights and impact the feasibility of achieving desired performance.

The overarching aspiration, science goals, science objectives, and Must / Discriminator requirements of the mission were developed during and after the June 2011 ExoPAG meeting with much input and discussion from the community. We drafted an initial mission statement, science goals, and science objectives and posted them for discussion to our community discussion board, the ExoPAG Flagship Mission Requirements SAG Yahoo Group. Two teleconferences were held in the summer of 2011 where the members of the SAG commented and iterated upon these drafts. Nearly 60 people (see Table 1) ultimately joined this effort. We reached consensus on these components of the report by August 2011, and then we drafted and refined the Musts and Discriminator Requirements from September through December. We reported on these efforts and the resulting body of work at the January 2012 ExoPAG meeting where this process and product was endorsed.

*Table 1: List of SAG participants*

| Daniel | Apai | Jeremy | Kasdin | Jagmit | Sandhu |
|---|---|---|---|---|---|
| Jean-Charles | Augereau | James | Kasting | Gene | Serabyn |
| Rus | Belikov | John | Krist | Stuart | Shaklan |
| Jeff | Booth | Marie | Levine | Michael | Shao |
| Jim | Breckinridge | Chuck | Lillie | Erin | Smith |
| Kerri | Cahoy | Doug | Lisman | Arif | Solmaz |
| Webster | Cash | Carey | Lisse | Rémi | Soummer |
| Joseph | Catanzarite | Amy | Lo | Bill | Sparks |
| Supriya | Chakrabarti | Rick | Lyon | Karl | Stapelfeldt |
| Mark | Clampin | Avi | Mandell | Angelle | Tanner |
| Denis | Defrere | Joe | Marley | Domenick | Tenerelli |
| Michael | Devirian | Mark | Marley | Wesley | Traub |
| Tiffany | Glassman | Michael | McElwain | John | Trauger |
| Tom | Greene | Charley | Noecker | Zlatan | Tsvetanov |
| Olivier | Guyon | Pascal | Petit | Maggie | Turnbull |
| halleyguy | | Joe | Pitman | Steve | Unwin |
| Sally | Heap | Marc | Postman | Robert | Vanderbei |
| Douglas | Hudgins | David | Redding | Amir | Vosteen |
| Lisa | Kaltenegger | Aki | Roberge | Darren | Williams |

## 2  Science Goals

The primary scientific goal of the exoplanet flagship mission is detecting and spectroscopically characterizing at least one Earth-sized planet in the habitable zone of a nearby Sun-like star. We have also expressed this in a **Mission Statement** for a broad, non-specialist audience:

> This mission will find potentially habitable planets and planetary systems orbiting nearby stars.

The mission's more specific **Science Goals** are:

**Goal 1:** Determine the overall architectures of a sample of nearby planetary systems. This includes determining the numbers, brightnesses, locations, and orbits of terrestrial to giant planets and characterizing exozodiacal dust structures in regions from habitable zones to ice lines and beyond. This information will also provide clues to the formation and evolution of these planetary systems.

**Goal 2:** Determine or constrain the atmospheric compositions of discovered planets, from giants down to terrestrial planets. Assess habitability of some terrestrial planets, including searching for spectral signatures of molecules and chemical disequilibrium consistent with the presence of life. Determining or constraining surface compositions of terrestrial planets is desirable but is not strictly required.

**Goal 3:** Determining or constraining planetary radii and masses are stretch goals of this mission. These are not strictly required. However, measuring radii and masses would provide a better understanding of detected planets, significantly increasing the scientific impact of this mission.

# 3   Science Objectives

These **Science Goals** are now broken down into **Objectives** that serve as the basis for the mission's exoplanet systems requirements.

**Objective 1:** Directly detect terrestrial planets that exist within the habitable zones around nearby stars or, alternatively, observe a large enough sample of nearby systems to show with high confidence that terrestrial planets are not present.

**Objective 2:** Measure or constrain orbital parameters (semi-major axis and eccentricity) for as many discovered planets as possible, especially those that show evidence of habitability.

**Objective 3:** Obtain absolute photometry in at least three broad spectral bands for the majority of detected planets. This information can eventually be used, in conjunction with orbital distance and planet radius, to constrain planetary albedos.

**Objective 4:** Distinguish among different types of planets, and between planets and other objects, through relative motion and broadband measurements of planet color.

**Objective 5:** Determining or constraining planetary masses is highly desired but not required. Determining masses would allow estimates of planetary radii to be made, thereby enabling calculation of planetary albedos (Objective 3).

**Objective 6:** Characterize at least some detected terrestrial planets spectroscopically, searching for absorption caused by $O_2$, $O_3$, $H_2O$, and possibly $CO_2$ and $CH_4$. Distinguish between Jupiter-like and $H_2O$-dominated atmospheres of any super-Earth planets. Such information may provide evidence of habitability and even of life itself. Search for Rayleigh scattering to constrain surface pressure.

**Objective 7:** Directly detect giant planets of Neptune's size or larger and having Jupiter's albedo in systems searched for terrestrial planets. Giants should be detectable within the habitable zone and out to a radius of at least 3 times the outer radius of the habitable zone.

**Objective 8:** Characterize some detected giant planets spectroscopically, searching for the absorption features of $CH_4$ and $H_2O$. Distinguish between ice and gas giants, as well as between Jupiter-like and $H_2O$-dominated atmospheres of any mini-Neptune planets.

**Objective 9:** Measure the location, density, and extent of dust particles around nearby stars in order to identify planetesimal belts and understand delivery of volatiles to inner solar systems.

**Objective 10:** In dusty systems, detect and measure substructures within dusty debris that can be used to infer the presence of unseen planets.

**Objective 11:** Understand the time evolution of circumstellar disk properties around a wider star sample at greater distances, from early protoplanetary stages through mature main sequence debris disks.

*Discussion*

These Objectives are not prioritized, but represent three broad categories of science investigations which can be done with the same observatory, and can energize the community around it.

Objective 1 specifically calls for a survey to detect Earth-like planets with this mission, rather than relying on prior detections by ground or space observatories. This is because the principal methods of prior detection (radial velocity, astrometry, transits, IR nulling interferometry) become either very challenging or very incomplete for the targets we're focusing on—Earth size planets in habitable zone orbits around nearby F, G, and K stars. Based on repeated analyses of these methods and a variety of instrument concepts, a consensus persists that a direct-imaging mission designed to *characterize* Earth-size planets in habitable zone orbits is also the best method for *finding* them first. Savransky et. al.[7] showed that prior astrometric knowledge of the presence and even the orbits of exoplanets would yield only modest improvement in the scientific productivity of a direct detection mission. These judgments have been durable for several years, but deserve continual reexamination. We should emphasize that these other measurements, whether before, during, or after the flagship mission, could substantially enrich the science harvest for exoplanets of all sizes, and NASA should continue to support them enthusiastically.

Two examples illustrate the difficulty of prior detection of Earth-like planets. The recent RV detection of α Cen Bb, with an amplitude of 51 cm/sec at 3.2 day period, is a benchmark for the effort needed. To conduct a survey of nearby stars, we would need sensitivity to amplitudes of 3-20 cm/sec with 60-2000 day periods, on stars which are as much as 6 magnitudes fainter than α Cen B. Similarly, a search proposed for the SIM space astrometry mission would have used 12,000 total observations of the 60 key stars to reach a total-mission detection sensitivity approaching 1 picoradian. These very challenging sensitivities needed for radial velocity or astrometric detection of Earth-like planets remain a significant technology hurdle, and so far, no method or program has been developed and shown to be capable of achieving them. Thus it's doubtful that inserting such a program before the flagship direct detection mission would be worth the cost and delay.

The above Science Goals and Objectives are related as follows:

| Science Goals | Science Objectives | | | | | | | | | | |
|---|---|---|---|---|---|---|---|---|---|---|---|
| | 1 | 2 | 3 | 4 | 5 | 6 | 7 | 8 | 9 | 10 | 11 |
| 1. Architectures | ✓ | ✓ | | ✓ | ✓ | | ✓ | | ✓ | ✓ | ✓ |
| 2. Compositions | | | ✓ | ✓ | (✓) | ✓ | | ✓ | | | |
| 3. Masses & radii | | | ✓ | ✓ | ✓ | | | | | ✓ | |

Note that every row and column has at least one checkmark.

# 4  Level 1 Requirements

We have determined preliminary requirements from these objectives, but finalizing some requirements will require better knowledge than is currently available of the frequency of Earth-like planets ($\eta_\oplus$, called eta_Earth) and the amount and distribution of exozodiacal dust. That said, we next present the preliminary, provisional requirements based on our current assumptions for these values.

Since we are preparing to recommend a mission architecture from among several competing options, these requirements are posed in a form that serves the decision process but is different from the traditional structure (minimum/baseline/goal requirements). Specifically,

a. What we have traditionally called the minimum mission requirements—below which the mission has insufficient scientific merit and should be canceled—are herein called "Must" requirements.

b. In place of baseline and goal (stretch) mission requirements, we list a number of "Discriminators," each of which is a criterion that represents added value in the science harvest.

If there is a minimum acceptable value of any Discriminator, it is included among the Musts; thus parallel language appears often in these two requirements sections. This decision process allows candidate missions to be compared on a variety of scientific, technical, and programmatic criteria even if they aren't comparable in cost and capability and have very different areas of excellence. The science-driven Musts and Discriminators are presented next.

# 5 Requirements

This list is primarily based on the TPF-C STDT requirements, translated into the new Musts/Discriminators form, which is described below. This form is preferred to a traditional requirements language because we will need to select a mission concept from among candidates with very different strengths and maybe cost. The present exercise should be viewed as a step in preparing for that complex decision. A traditional set of requirements has typically tended to bias the selection by emphasizing one criterion over others. A rough analogy is making an object that must fit inside a wooden box vs. one that fits inside a bag; the size of the bag allows comparisons between objects of very different shape and dimensions, without over-emphasizing the specific shape.

## *5.1 Assumptions and Definitions*

On the whole we will stick with the definitions in Sec 1.2 of the STDT report. They are echoed here, in some cases with a slightly different flavor.

| EID | Equivalent Insolation Distance; i.e. the distance between the star and planet for which the stellar irradiance is equal to that in our own solar system at a specified distance. For example, at 1 AU EID in the exoplanet system, the irradiance is the same as that here on Earth, even though the true distance is larger or smaller because of the star luminosity. |
|---|---|
| HZ | Habitable Zone, extends from 0.75-1.8 AU (EID) |
| IHZ | Inner HZ, extends from 0.7-1.0 AU (EID) |
| CumHZ | Cumulative partial Habitable Zones, the sum of the fraction of the HZ observed on each star during the mission. This excludes repeat observations of the same regions of orbital period, orientation, and phase. |
| CumIHZ | Cumulative partial INNER Habitable Zones, the sum of the fraction of the IHZ observed on each star over the entire mission. Note the distinction between the entire HZ and just the IHZ. |
| SMA | Semi-major axis, half the diameter of the long axis of an elliptical orbit. |
| TXP | Terrestrial eXoPlanet: defined as 0.8-2.2 $R_{earth}$, with SMA in the HZ, and eccentricity <0.2. We also adopt these assumptions: assume $dN/da \propto a$, and $dN/dM \propto 1/M$, uniform eccentricity distribution, with geometric albedo of 0.2 in the full science passband. |
| Candidate exoplanet | Point source in region of interest with appropriate brightness relative to the star. |
| Confirmed exoplanet | Shows common proper motion or recognizable high-res spectrum. Able to distinguish between planets and background confusion sources, and exozodiacal dust structures. Confusion can be broken using broadband colors, spatial resolution, spectra, or proper motion, whatever works most efficiently, high spatial resolution, spectra, possibly broadband colors or changing brightness with phase. |
| Kuiper belt | Debris belt at >10 AU with surface brightness >24 mag/arcsec$^2$. |
| HZ exozodi | Exozodi surface brightness in habitable zone of 10× (TBR) that of a solar system twin at median inclination, with no asymmetries. LBTI observations are expected to reach this sensitivity,[8] so we should have statistically significant exozodi brightness data to this level. We assume every system is as bright as this measurement limit. |
| Confusion sources | Assume no confusion sources in the FOV. Discrimination from confusion sources is an important problem to address, but our knowledge is insufficient at this time. |
| IWA | Inner Working Angle. The minimum angular separation from the central star at which a faint point source has at least 50% throughput. |
| OWA | Outer Working Angle. The maximum angular separation from the central star at which detection of a faint point source requires an integration time no more than 4× (TBR) that of an object of the same brightness at the best location within 0.5" of the star. For some star-suppression systems, the integration time rises sharply beyond some angular radius, the Nyquist angle given by the deformable mirror size. |
| δ-mag | The brightness ratio given in magnitudes between the central star and a faint point source that can be detected with high confidence. This can vary with angle from the star |
| SNR | Signal to noise ratio |
| FAP | False alarm probability, the probability that a point source that appears to be a planet would turn out to be something else. |
| "Detect" a planet | SNR compatible with FAPs of 1% (TBR). There should be a FAP for the planet search, another for confirmed exoplanets, and another for fully characterized exoplanets |
| TBR/TBD | To be revised/ To be determined |

## *5.2 Musts*

The following are pass/fail bare minimum requirements for the mission to be considered worthy of the effort and expense. All candidate mission concepts must meet these criteria.

M1  Able to detect an Earth twin at quadrature in a Solar System twin at a distance of 10 pc
*Rationale:* "Pushpin" in the middle of the performance range required by M3. That is, any observatory able to meet M3 should naturally meet this as well.
*Comment:* Not a driving requirement, but helpful to communicate with NASA and taxpayers.
*Maps to*: O1

M2  Able to detect a Jupiter twin at quadrature in a Solar System twin at a distance of 10 pc
*Rationale:* "Pushpin" in the middle of the performance range required by M3.
*Comment:* Not a driving requirement, but helpful to communicate with NASA and taxpayers.
*Maps to:* O7

**M3** Examine at least 14 CumHZs to detect point sources with TXP sensitivity
*Rationale:* Matches the STDT's Requirement 3 for a minimum mission (§1.4.2), with optimistic $\eta_\oplus$=20%. We chose this case for the Musts, so that we can still consider a mission smaller than the classic TPF-C. This case also yields >95% probability of seeing at least one TXP assuming $\eta_\oplus$ =20%, and also offers a good chance of seeing several TXPs.
NB: the IWA and $\delta$-mag needed to satisfy M3 are also sufficient to detect many giant planets outside the HZ.
*Comment:* If $\eta_\oplus$ =20%, the expected value of the number of TXPs detected is 2.8. The probability of seeing *at least one* TXP can be estimated by
$$P(1;CumHZ, \eta_\oplus) = 1 - P(0;CumHZ, \eta_\oplus) = 1-(1-\eta_\oplus)^{CumHZ} = 1-0.8^{14} = 95.6\%$$
Note that our "optimistic" $\eta_\oplus$ is supported by a preliminary analysis of the Kepler data[9], which argues for a value of more than 30%.
*Maps to:* O1, O7

**M4** Examine at least 3 (TBR) CumIHZs to detect point sources with TXP sensitivity
*Rationale:* We want to ensure that not all of the partial HZs examined are in the outer HZ, 1-2 AU (EID). As with M3, this establishes capabilities that allow giant outer planet detection.
*Comment:* 3 was chosen semi-arbitrarily; this warrants more thought, and a capability assessment. At least we would like this number of CumIHZs to be naturally consistent with the capability of a mission that is sized to meet M3 above, assuming a reasonable distribution of SMA within the HZ.
*Maps to:* O1, O7

**M5** Characterize every discovered candidate exoplanet by R>=4 spectroscopy (color photometry) across a passband from 0.5 µm to the maximum feasible wavelength less than 1.0 µm.
*Rationale:* Require <u>instrumentation</u> and <u>time allocation</u> to attempt this measurement on every planet found, large or small. Long wavelengths may be unreachable due to IWA or red leak.
*Comment:* Some are concerned that this "do whatever you can" language has no teeth. But others are concerned that alternative language will lead to impossible requirements.
*Maps to:* O3, O4, O8

**M6** Able to characterize the "Earth" in a Solar System twin at 5 pc (TBR) and the "Jupiter" in a Solar System twin at 10 pc by R>70 spectroscopy across 0.5-1.0µm
*Rationale:* Require instrumentation and enough observing time for one such measurement. Assume favorable conditions in which IWA and brightness are not a limitation. The second clause about Jupiter connects a Must to O8, but we expect the mission to meet this easily.
*Comment:* Pushpin for hypothetical optimistic case. Not all found planets will be reachable by spectroscopy to 1.0µm because of IWA limitations; but if IWA scales with $\lambda$, then detection at 10 pc at $\lambda$=0.5µm is equivalent to 5 pc at $\lambda$=1.0µm. Similarly,

| (10 pc) · (0.5µ) / (0.94µ) = | 5.3 pc | $H_2O$ |
| (10 pc) · (0.5µ) / (0.76µ) = | 6.6 pc | $O_2$ |

The 10 pc distance chosen for Jupiter is fairly arbitrary, not challenging in photometry or IWA. Its purpose was just to make a requirement for outer giant planet spectroscopy. Also note that for some mission concepts, IWA is approximately independent of wavelength across a wide range.
*Maps to:* O6, O8

**M7**   Able to determine the orbital SMA to 10% for the "Earth" in a Solar System twin at 6.5 pc
*Rationale:* Like in STDT §1.4.2 (4)
*Comment:* Pushpin for hypothetical optimistic case. We declare that this knowledge has value, but our intent at this time is that IWA will not be the main challenge; it just requires instrumentation for star-planet angle measurements, and an adequate observing strategy. The 6.5 pc distance is fairly arbitrary in meeting that intent.
*Maps to:* O1, O2, O4

**M8**   Able to measure $O_2$ A-band equivalent width to 20% for the "Earth" at elongation in a Solar System twin at 6 pc.
*Rationale:* Establish measurement sensitivity to a key biomarker spectroscopic signature.
*Comment:* If IWA scales with $\lambda$, and the planet can be detected at 10 pc at $\lambda=0.5\mu m$, then it can be detected at 6 pc at $\lambda=0.83\mu m$, which is sufficient to span the $O_2$ A-band at $\lambda=0.76\mu m$.
*Maps to:* O6

**M9**   Able to measure $H_2O$ equivalent width to 20% for the "Earth" at elongation in a Solar System twin at 5 pc and the $CH_4$ equivalent width in a "Jupiter" in a Solar System twin at 10 pc.
*Rationale:* Establish measurement sensitivity to a key biomarker spectroscopic signature. Was not included in STDT §1.4.2, but it could be assuming IWA scales proportional to $\lambda$.
*Comment:* If IWA scales with $\lambda$, and the planet can be detected at 10 pc at $\lambda=0.5\mu m$, then it can be detected at 5 pc at $\lambda=1\mu m$, which is sufficient to span the $H_2O$ band at $0.94\mu m$. Likewise, there is a strong $CH_4$ band at $0.889\ \mu m$, which we expect to be accessible at 0.5" working angle.
*Maps to:* O6, O8

**M10**  Conduct a search that has at least 85% (TBR) probability of finding at least one TXP <u>and</u> measuring its color at R=4 <u>and</u> measuring its SMA with 15% uncertainty (TBR) <u>and</u> measuring its spectrum (0.5-0.8µm)(TBR) with R≥70 and 20% (TBR) spectrophotometric uncertainty.
*Rationale:* The combination of several key measurements on one planet. This is full of TBRs, which will require a lengthy analysis to resolve; but it illustrates a tasty minimum likelihood of finding and coarsely-but-fully characterizing a TXP. This implicitly constrains search depth, time allocation, and characterization capability.
*Comment:* This is much more difficult than M3—being able to measure color, SMA, fine spectrum to 0.8µm, and 20% photometry **all on the same TXP.** If we don't scale back the parameters in this case, the observatory will be driven strongly by this requirement, and likely go well beyond the other requirements. We still don't know that a planet exists with characteristics that are favorable for all of these measurements together, so we can't assemble requirements that <u>will get</u> that one planet; but again we can substitute probabilities for the scientific unknowns ($\eta_\oplus$ and orbit/IWA), and then estimate the statistical likelihood of it for any mission concept.
*Maps to:* O1, O2, O3, O4, O6

**M11**  Absolute photometry of "Earth" at maximum elongation in a Solar System twin at 8 pc to 10%
*Rationale:* Like in STDT §1.4.2 (6), which refers to an Earth twin in a Solar System twin at 8 pc. Pushpin to fix a calibration requirement
*Maps to:* O3

M12    Able to guide on the central star as faint as $V_{AB}$=16 (TBR) for high contrast imaging at degraded sensitivity.
*Rationale:* Contrast for disk science is not as demanding as for TXP science, but generally demands a wider range of stars, often much fainter than TXP target stars.
*Comment:* We need further conversation with the SAG 1 team (characterization of exozodi disks). We hope this will also prompt a capability assessment. We are hoping for graceful degradation of coronagraphy with central star magnitude. A goal is sensitivity to mag 30 point sources in the neighborhood of a star of any magnitude.
*Maps to:* O9, O10, O11

M13    Capable of high-contrast optical imaging of extended structures with surface brightness sensitivity of (TBD of the star) at > TBD arcsec from the central star.
*Rationale:* Disk science
*Comment:* We need further conversation with the SAG 1 team (characterization of exozodi disks). Probably need a few such benchmarks on a curve. The precise values of these parameters cannot be determined until a sensitive, high angular resolution study of the exozodiacal dust of nearby stars is completed.[8]
*Maps to:* O9, O10, O11

N.B. there are no Musts for a number or percentage of *confirmed* exoplanets. Confirmation is a knotty problem, not well understood, and it may prove too big a challenge for the first mission we can afford. We would still get a list of exoplanet *candidates* and a significant scientific and technical step forward. See the mapping of Musts to Objectives at the end of the next section.

### *5.3 Discriminators*

The following are Discriminators, which are not pass/fail but numerically scored based on quantitative or semi-quantitative metrics. The metrics are expected to be well-defined and unambiguous, like observatory mass, number of launch vehicles, number of science observations in 5 years, etc., and should be defined in a way that is applicable to all concepts.

The scores are rooted in those metrics and are ideally developed by consensus, but often fairly subjectively. Scores are a layer of abstraction from the metrics, to allow many Discriminators to be taken into consideration together, even though they may be of a dramatically different character. The set of Discriminators should be complete enough to allow each mission concept to accrue points for all of its strengths.

A set of weights are also developed by consensus, and reflect the relative importance of each Discriminator to the outcome of the mission. Each Discriminator has a numerical weight which applies to all concepts for that Discriminator. For each concept, a dot-product of the column of scores with these weights yields a single number, a composite score for the concept, which is the basis for choosing a mission concept. The scores and weights are both subjective, but we will conduct extensive tests of fiddling with these numbers to see how sensitive the final conclusion is to minor changes. If at the end we are comfortable that the decision rests on judgments that we all believe, we are ready to report a decision with confidence.

D1  Number of CumHZs searched to TXP sensitivity
*Rationale:* Beyond the minimum in M3, we want a deeper search (more CumHZs) to get more planets
*Comment:* An earlier version of this requirement specified a minimum δ-mag, but this was deemed redundant and overspecifying. We preferred staying close to (a) the probability of at least one planet and (b) the expected value of the number of planets.
*Maps to:* O1, O7

D2  Number of CumIHZs searched to TXP sensitivity
*Rationale:* Similarly, we want a deeper search of the IHZ; cf. M4 - more CumIHZs fills in the inner planets
*Maps to:* O1, O7

D3  Minimum brightness of exoplanet that is detectable at angles in the range from 1-2×IWA (TBR).
*Rationale:* Ability to see fainter point sources improves the depth of search (cf. M3, M4) and its completeness down to small sizes; also improves characterization by virtue of seeing more of the orbit. Typically δ-mag = 26, but larger δ-mag gets more planets.
*Maps to:* O1, O7

D4  Number of candidate exoplanets that are confirmed
*Rationale:* Establish the capability to do measurements to confirm candidate exoplanets.
*Comment:* See definition of "Confirmed." Confirming every exoplanet system could be very demanding for some mission concepts. Relaxing this number may leave many planet candidates unproven until a followup mission.
*Maps to:* O1, O7

D5  Number of discovered exoplanets characterized by R>4 spectroscopy (color photometry) across the full 0.5-1.0µm
*Rationale:* See M5. If there's any limitation or difficulty, it's better to characterize more planets by color.
*Maps to:* O3, O4, O8

D6  Number of discovered TXPs and giant planets that can be characterized by R>70 spectroscopy across the full 0.5-1.0 µm
*Rationale:* See M6. It's better to characterize more planets for the presence of $H_2O$, e.g. by having a small IWA. These capabilities also aid the characterization of giant planets outside the HZ.
*Comment:* Again, this is a statistical estimate based on distributions and observing scenarios.
*Maps to:* O6, O8

D7  Number of discovered TXPs characterized by R>70 spectroscopy across 0.5-0.85 µm
*Rationale:* See M7. It's better to characterize more planets by $O_2$ even if $H_2O$ is inaccessible. These capabilities also aid the characterization of giant planets outside the HZ, e.g. via methane at 728, 793, and 863nm, and water at 830 nm.
*Comment:* Again, statistical estimate based on distributions and observing scenarios.
*Maps to:* O6, O8

D8  Extended passbands to NIR and NUV
*Rationale:* Some mission concepts are capable of TXP sensitivity further into the IR or the UV. This can provide more atmospheric absorption bands and other information about the planet and exozodi.
*Maps to:* O6, O8

D9  Number (or percentage) of found candidate exoplanets for which we can test for common proper motion
*Rationale:* See D4 and the definition of "Confirmed." Common proper motion is the gold standard for proving the object is a true companion; some alternatives were listed above.
*Comment:* We don't know how many candidates will be detected, so we should not pin ourselves to a fixed *number*. And in an exoplanet-rich scenario, confirming a minimum *percentage* may be a challenge.
*Maps to:* O1, O7

D10  Number of found planets whose orbital SMA can be determined with ±10% uncertainty (TBR) or better.
*Rationale:* This may be difficult because of the number of visits required. This depends on agility for multiple revisits, confident detection each time, and accurate planet-star relative astrometry.
*Comment:* Perhaps also give credit for even finer SMA determination.
*Maps to:* O1, O2, O4

D11  Number of TXP masses determined to TBD%
*Rationale:* Measurement of the host star's astrometric wobble is the gold standard for exoplanet mass determination.[10] (Indirect methods have been proposed, but are vulnerable to scientific uncertainties.) No existing well-developed mission concepts are believed capable of providing this astrometric information, so there is no Must or minimum requirement for this knowledge. But if we can demonstrate convincingly that one or more concepts could provide this, we should give high scores for that.
*Maps to:* O4, O5

D12  Number of discovered TXPs characterized by absolute photometry
*Rationale:* See M10 – we want more planets characterized by absolute photometry
*Comment:* Again, statistical estimate based on distributions
*Maps to:* O3, O4

D13  Number of giant exoplanet candidates detected in ExoEarth target systems
*Rationale:* We want the capability to detect and characterize a variety of giant planets, especially to see if there are correlations between the presence and nature of TXPs and of giant planets. Also establishes the virtue of a large ratio OWA/IWA.
*Maps to:* O7, O8, O11

D14  Number of Kuiper Belts imaged in ExoEarth target systems
*Rationale:* Of course we want to detect many examples of inner and outer debris disks, but we especially want to see if there are correlations between the presence and nature of TXPs and of Kuiper Belts. Also establishes the virtue of a large ratio OWA/IWA.
*Comment:* We haven't defined "Kuiper Belt" by a range of characteristics.

*Maps to:* O9, O10, O11

### 5.4 Mapping of Musts and Discriminators to Objectives

Note that all rows in the following tables have at least one check mark. Also, all columns except O5 have at least one check mark in *each* table; O5 is captured in D11, and its absence from the Musts is explained in the rationale for D11.

|  | Science Objectives | | | | | | | | | | |
|---|---|---|---|---|---|---|---|---|---|---|---|
| **Musts** | 1 | 2 | 3 | 4 | 5 | 6 | 7 | 8 | 9 | 10 | 11 |
| M1: detect Earth twin | ✓ | | | | | | | | | | |
| M2: detect Jupiter twin | | | | | | | ✓ | | | | |
| M3: 14 CumHZs | ✓ | | | | | | ✓ | | | | |
| M4: 3 CumIHZs | ✓ | | | | | | ✓ | | | | |
| M5: colors | | | ✓ | ✓ | | | | ✓ | | | |
| M6: fine spectra | | | | | | ✓ | | ✓ | | | |
| M7: orbital SMA | ✓ | ✓ | | ✓ | | | | | | | |
| M8: oxygen | | | | | | ✓ | | | | | |
| M9: water | | | | | | ✓ | | ✓ | | | |
| M10: all on 1 planet | ✓ | ✓ | ✓ | ✓ | | ✓ | | | | | |
| M11: absol photometry | | | ✓ | | | | | | | | |
| M12: guide on faint star | | | | | | | | | ✓ | ✓ | ✓ |
| M13: surface brightness | | | | | | | | | ✓ | ✓ | ✓ |

| **Discriminators** | 1 | 2 | 3 | 4 | 5 | 6 | 7 | 8 | 9 | 10 | 11 |
|---|---|---|---|---|---|---|---|---|---|---|---|
| D1: # CumHZs | ✓ | | | | | | ✓ | | | | |
| D2: # CumIHZs | ✓ | | | | | | ✓ | | | | |
| D3: max δ-mag | ✓ | | | | | | ✓ | | | | |
| D4: # confirmed | ✓ | | | | | | ✓ | | | | |
| D5: # planets, 4 color | | | ✓ | ✓ | | | | ✓ | | | |
| D6: # planets, full spectra | | | | | | ✓ | | ✓ | | | |
| D7: # planets, part spectra | | | | | | ✓ | | ✓ | | | |
| D8: NIR and NUV | | | | | | ✓ | | ✓ | | | |
| D9: common PM | ✓ | | | | | | ✓ | | | | |
| D10: # orbit SMA | ✓ | ✓ | | ✓ | | | | | | | |
| D11: # astrometric mass | | | | ✓ | ✓ | | | | | | |
| D12: # absol photometry | | | ✓ | ✓ | | | | | | | |
| D13: # giants w/ TXPs | | | | | | | ✓ | ✓ | | | ✓ |
| D14: # KuiperB w/ TXPs | | | | | | | | | ✓ | ✓ | ✓ |

# 6 Conclusion

We believe this captures most of the features we value in a flagship mission concept, and prepares a process for selecting the best one. But of course, we expect modifications and additions to this list as our understanding improves.

More importantly, recent programmatic developments have motivated a look at how a smaller mission (1.5-2.5m telescope) might achieve some of these objectives in the near term, in lieu of a flagship which might take much longer. Indeed, a particular new 2.4m telescope opportunity seems like a possible path, but it would require great flexibility and compatibility with other astronomy objectives. This will make the selection of a planet-finding method even more exciting and more complicated at the same time.

## Acknowledgments

We are grateful for Marie Levine's assistance in facilitating this effort and keeping momentum. Some support for this research was provided by the Jet Propulsion Laboratory, California Institute of Technology, under a contract with the National Aeronautics and Space Administration.